\documentclass[11pt, onecolumn, a4paper]{article}
\usepackage{amsmath}
\usepackage{amssymb}
\usepackage{amsthm}
\usepackage{lipsum}
\usepackage{mathptmx}

\usepackage[rm,up,sc,topmarks,calcwidth,pagestyles]{titlesec}
\titleformat{\section}{\Large\bfseries\rmfamily}{\thesection}{1em}{}
\titleformat{\subsection}{\large\bfseries\rmfamily}{\thesubsection}{1em}{}
\titleformat{\subsubsection}{\large\it\rmfamily}{\thesubsubsection}{1em}{}

\usepackage{cite}
\usepackage[dvipdfmx]{graphicx}
\usepackage{textcomp}
\usepackage{url}

\usepackage{color}
\usepackage{algorithm}
\usepackage{algorithmic}
\title{Analytics of Adaptive Online Testing in Practice Over a Decade}
\author{Hideo Hirose\footnote{Chuo University \& Kurume University, Japan} }
\date{}
\begin{document}
\maketitle
\thispagestyle{empty}

\section*{Abstract}
Adaptive online testing efficiently assesses examinee proficiency by dynamically adjusting the difficulty of test items based on their performance. To achieve this, items are selected so that their difficulty closely matches the test taker's estimated ability at each stage of the test. This alignment implies that the probability of a correct answer tends toward 0.5. However, in practical settings, this probability may not converge to 0.5 unless the test comprises a sufficiently large number of items. This could give the impression that the adaptive mechanism is not functioning properly. 
Nevertheless, even when the number of items is small, such as 5 or 7, the adaptive nature of the system can still be observed by examining the relationship between item difficulty and the mean estimated ability of examinees for the corresponding item. Considering that each item typically requires about 2-5 minutes to solve, this number of items appears to be a practical choice that avoids excessive fatigue for test-takers.
\\[2mm]
{\it Keywords: } learning analytics, computer adaptive testing, correct answer rate, item response theory, multiple-choice problem, practical number of items.

\section{Introduction}

Suppose a test-taker with ability $\theta$ attempts a question with difficulty $b$. For simplicity, assume that the probability $p$ of successfully solving the problem follows the standard normal distribution $N(0,1)$. Then, if $\theta < b$, we expect $p < 0.5$; if $\theta > b$, then $p > 0.5$; and if $\theta = b$, then $p = 0.5$.

Adaptive testing systems aim to select questions whose difficulty levels are closely aligned with the examinee's estimated ability, in order to obtain a reliable estimate using as few items as possible. This is because, from a statistical perspective, the largest information is gained when the difficulty of an item matches the test-taker’s ability level~\cite{Lord}.
Consequently, in adaptive testing, as each question is selected to match the test-taker’s estimated ability, the correct answer rate is expected to converge asymptotically to 0.5.
 In this paper, the correct answer rate (CAR)~\cite{PISM2018} is defined as the ratio of the total number of correct answers to the total number of attempted questions. In online testing systems, some test-takers may discontinue answering a question before completing it. Such cases are also included in the total number of responses when calculating the CAR in this study.

\begin{figure}[htbp]
\begin{center}
\includegraphics[height=9.5cm]{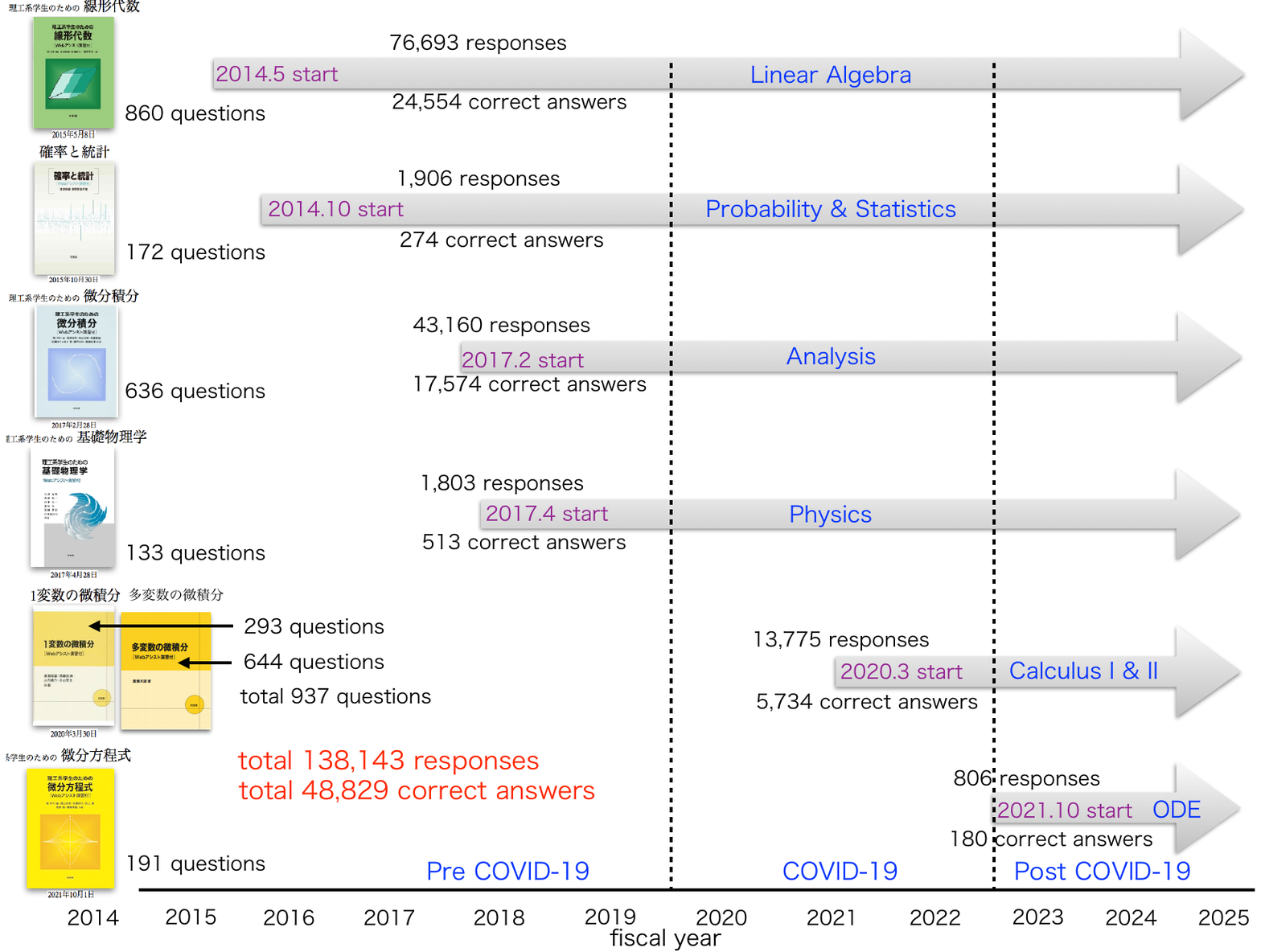}
\caption{Adaptive online testing system bundled with undergraduate mathematics textbooks. The numbers of questions, total responses, and total correct answers accumulated over more than 10 years are shown.}
\label{figs:ResponsesToSubjects}
\end{center}
\end{figure}

\medskip
However, regarding the CAR, an analysis of an adaptive online testing system for undergraduate mathematics, implemented and refined over more than a decade, reveals a different perspective, as will be discussed below. 
The system was initially developed in 2013 and first introduced for Linear Algebra courses in 2014~\cite{LTLE2016c, PISM2018}. Since then, as shown in Figure~\ref{figs:ResponsesToSubjects}, the system has expanded to cover additional subjects such as Probability and Statistics, Analysis, Physics, Calculus I and II, and Ordinary Differential Equations (ODE). In the figure, the book covers and the launch dates for each subject are displayed, indicating that the online questions are integrated with the respective textbooks.
Since its initial deployment in 2014, the system has been in continuous operation. Even during the COVID-19 pandemic from 2020 to 2022 (see Appendix~1), the service was maintained without interruption.

Figure~\ref{figs:ResponsesToSubjects} also presents the total number of responses and correct answers for each subject. The overall number of responses has reached 138,143, among which 48,829 were correct. Accordingly, the overall CAR is 0.353 (95\% confidence interval: [0.351, 0.356]), which is noticeably lower than the expected value of 0.5. 
Table~\ref{table:ActualCorrectAnswerRates} shows the actual CARs for each subject, all of which are below 0.5.

\begin{table}[htbp]
\caption{Actual correct answer rate, CAR, to each subjects.
\label{table:ActualCorrectAnswerRates}}
\begin{center}
\begin{tabular}{c|c|c}
\hline
subject & correct answer rate, CAR & $95\%$CI$^*$\\
\hline
Linear Algebra & 0.320 & [0.317, 0.324]\\
Probability \& Statics & 0.144 & [0.128, 0.160]\\
Analysis & 0.407 & [0.402, 0.412]\\
Physics & 0.285 & [0.263, 0.306]\\
Calculus I \& II & 0.416 & [0.408, 0.425]\\
ODE & 0.223 & [0.194, 0.253]\\
\hline
\end{tabular}
\center{CI$^*$: confidence interval}
\end{center}
\end{table}

\medskip
When asymptotic theory is taken into account, one might mistakenly equate theoretical predictions with actual outcomes. However, empirical observations indicate that there exists a substantial discrepancy between theoretical expectations and observed values when the number of items in the adaptive test sequence is relatively small.
The primary objective of this study is to quantify the extent of this discrepancy in scenarios where the number of consecutively administered items in an adaptive test is limited. 
To the best of our knowledge, no existing studies have utilized large-scale real-world data to systematically compare empirical correct answer rates (CAR) with those derived from theoretical models in the context of adaptive online testing. This paper presents the results of such an analysis.

\section{A brief survey on adaptive online testing}

Regarding the achievement of a CAR of 0.5, Bickel et al.~\cite{Bickel2001} discussed the maximization of item information and the alignment of item difficulty with examinee ability, a topic also addressed by Lord~\cite{Lord}.
Metsamuuronen~\cite{Metsamuuronen2023} pointed out that observed correct answer rates can be biased estimators of underlying item difficulty and proposed bias-corrected item difficulty measures.
Antal et al.~\cite{Frey2024} compared several item selection strategies to mitigate item exposure issues in online testing systems.
Bartroff, Finkelman, and Lai~\cite{BartroffFinkelmanLai2008} extended asymptotic optimality theory to computerized adaptive sequential testing.
However, there are few representative references specifically addressing the value of CAR.

From a psychological perspective, Ling et al.~\cite{Ling2017} demonstrated that simpler adaptive tests result in higher engagement and lower anxiety compared to both adaptive and fixed-item tests. Yao~\cite{Yao2019} used simulations to compare computer adaptive testing models for sentence item selection with passages.
Luo and Yang~\cite{LuoYang2024} showed that a cognitive design system approach offers solutions through automatic item generation.
Although psychological factors are often overlooked, they are important for understanding test takers’ motivation to engage with the exam.

To avoid the costly process of collecting large-scale, dense item response data and training diagnostic models, while addressing selection bias, Kwon et al.~\cite{Kwon2023} proposed the user-wise aggregate influence function method. Liu et al.~\cite{Liu2024} conducted a survey on computer adaptive testing focusing on machine learning aspects. Meta-analysis techniques have also been applied to adaptive online testing~\cite{Frey2024}.
These contributions appear useful from the standpoint of practical system operation.

Nevertheless, very few studies have investigated the functionality of adaptive systems based on empirical data from actual online exercises conducted by undergraduate students at established universities over periods exceeding ten years.

This paper presents findings derived from a substantial amount of such empirical data.
Moreover, the system examined here is uniquely designed to complement university mathematics textbooks with an online practice exercise system, which constitutes a distinctive feature of this work.

\section{Ability Estimation using Adaptive Online System}

\subsection{Common item response theory}

Since our adaptive online testing system is based on Item Response Theory (IRT)~\cite{Ayala, Hambleton91, LindenHambleton, Linden}, we first provide a brief overview of conventional IRT before describing our adaptive testing framework.

\begin{figure}[htbp]
\begin{center}
\includegraphics[height=6cm]{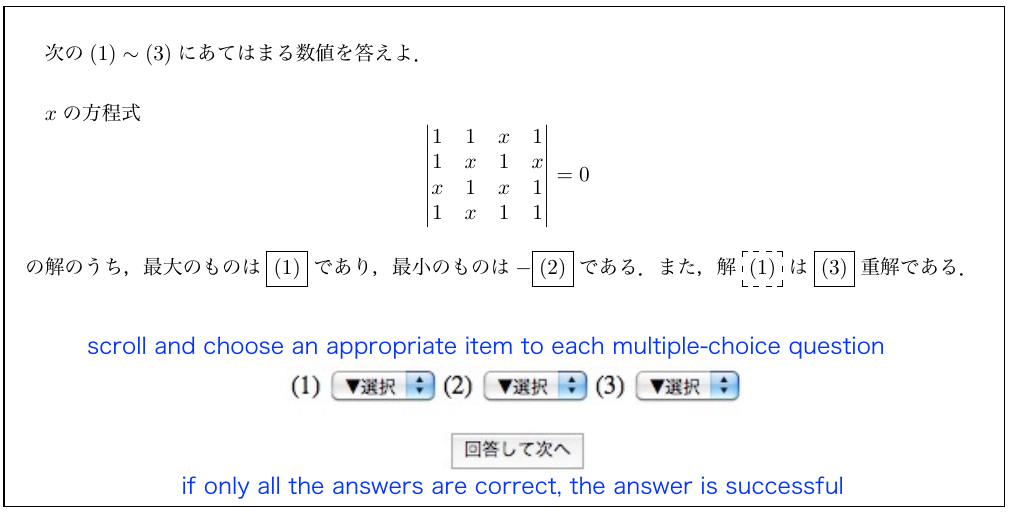}
\caption{An example of the multiple-choice type problem. To one multiple-choice question, an integer number including $0$ can be selected by scrolling. In this problem, since three multiple-choice questions should be answered, the probability of correct answer is $0.01$ without knowledge.}
\label{figs:questionexample}
\end{center}
\end{figure}

Consider a scenario where $n$ examinees simultaneously take a test consisting of $m$ items. That is, all examinees are administered the same set of questions.
In standard IRT, both the abilities of the $n$ examinees and the parameters of the $m$ items are estimated simultaneously using maximum likelihood estimation. In this study, we employ the two-parameter logistic model, defined by the item characteristic curve:
\begin{eqnarray} 
\label{eq:logisticfunction}
P_{ij}=P(\theta_i;a_j,b_j)={1 \over 1+\exp\{-1.7a_j(\theta_i-b_j)\} }=1-Q_{ij},
\end{eqnarray}
where $\theta_i$ represents the ability of examinee $i$, and $a_j$ and $b_j$ are the discrimination and difficulty parameters for item $j$, respectively. The constant 1.7 is included to make the logistic function closely approximate the standard normal distribution $N(0, 1)$.
We adopt the two-parameter logistic model rather than the three-parameter version (which includes a pseudo-guessing parameter) because each problem in our system typically comprises several multiple-choice questions, as illustrated in Figure~\ref{figs:questionexample}.

The likelihood function for the observed responses is given by:
\begin{eqnarray} 
\label{eq:likelihoodfunction}
L=\prod_{i=1}^n \prod_{j=1}^m \left(P_{ij}^{\delta_{ij}} 
\times Q_{ij}^{1-\delta_{ij}} \right),
\end{eqnarray}
where $\delta_{ij}$ is an indicator variable such that $\delta_{ij} = 1$ for a correct answer and $\delta_{ij} = 0$ otherwise. The model parameters $\theta_i$, $a_j$, and $b_j$ are estimated by maximizing this likelihood function.
Several estimation methods are available for this purpose, including Expected A Posteriori (EAP), Maximum A Posteriori (MAP), and Markov Chain Monte Carlo (MCMC) approaches~\cite{Baker, CATE}.
For further details on our IRT applications, refer to~\cite{EducationReview, Infomation2014, ComputerEducation}.

\subsection{Adaptive online testing system}

\begin{figure}[htbp]
\begin{center}
\includegraphics[height=4.5cm]{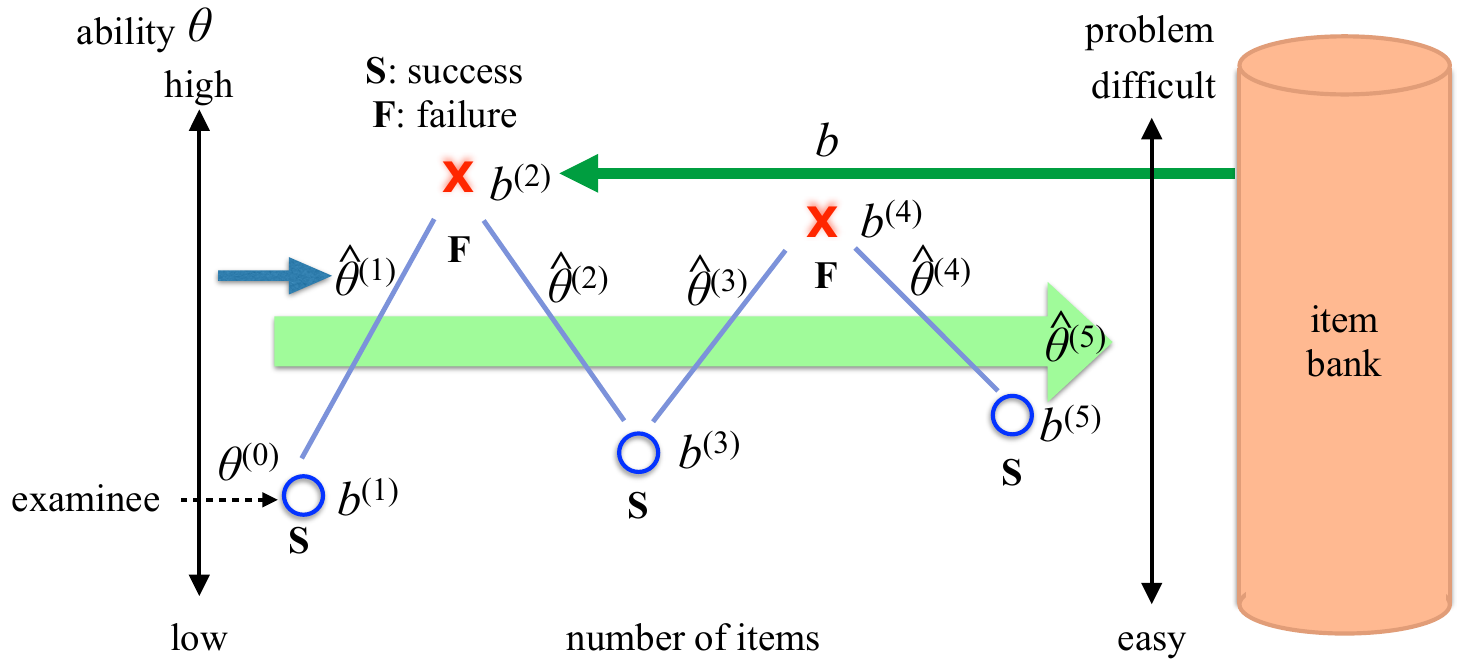}
\caption{Typical adaptive test procedure. Appropriate questions are presented consecutively to an examinee by the adaptive system using an item bank. After each answer, the system provides an updated estimate of the examinee’s ability at that point.}
\label{figs:adaptivetest}
\end{center}
\end{figure}

\begin{figure}[htbp]
\begin{center}
\includegraphics[height=6cm]{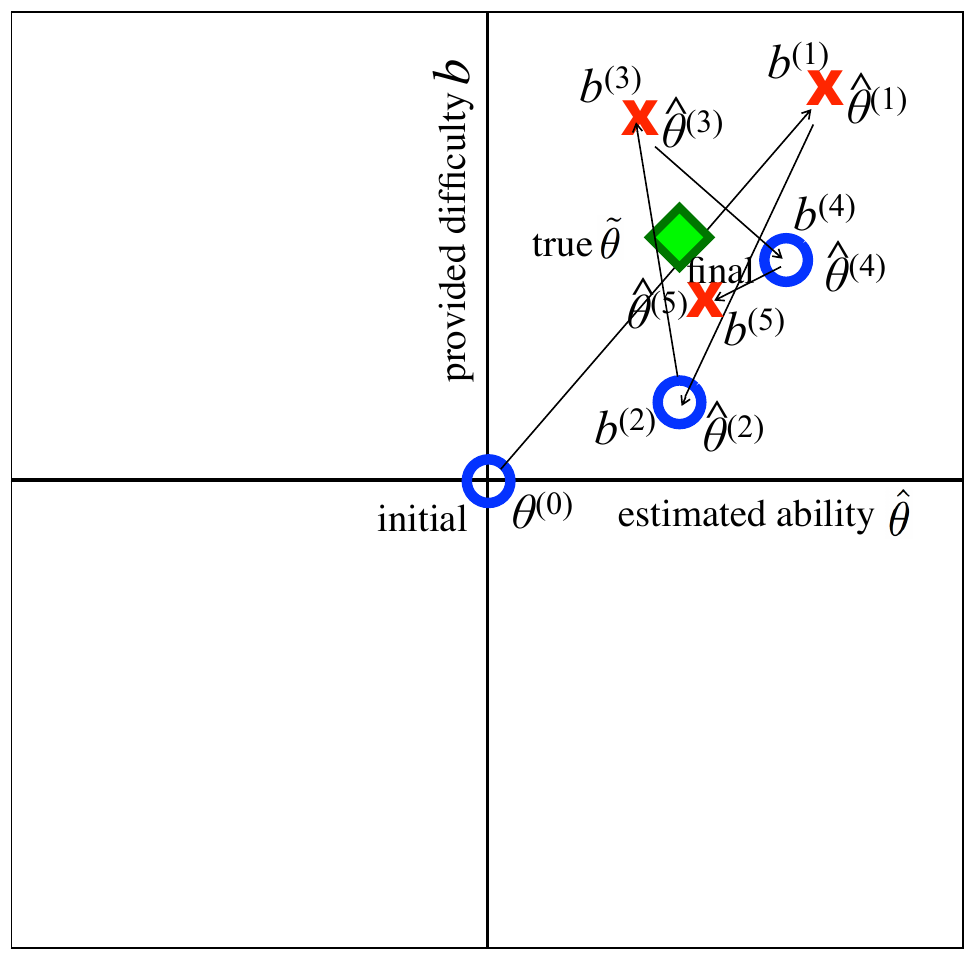}
\caption{Relationship between item difficulty and estimated ability during the adaptive testing process.}
\label{figs:process}
\end{center}
\end{figure}

However, in adaptive testing systems, each examinee takes a test composed of the most appropriate items presented sequentially.
Figure~\ref{figs:adaptivetest} illustrates a typical adaptive test procedure. If the examinee answers the first question correctly, the system selects a more difficult item for the next step. Conversely, if the response is incorrect, an easier item is administered. This up-and-down adaptive procedure continues until the predetermined number of items has been administered; in the present example, the test consists of five items. Such a mechanism is conceptually similar to the stress-strength model~\cite{SSmodel}.
Figure~\ref{figs:process} shows the relationship between the sequentially estimated ability values and the difficulties of the corresponding items. In the figures, $\hat{\theta}^{(1)}, \hat{\theta}^{(2)}, \cdots, \hat{\theta}^{(5)}$ denote the sequential ability estimates for the examinee, while ${b^{(1)}}, {b^{(2)}}, \cdots, {b^{(5)}}$ represent the difficulties of the items administered in order.

To select the most appropriate item at each step in accordance with the examinee’s proficiency level, it is assumed that all item parameters are known in advance. In practice, however, these parameters are not always available at the start of the adaptive testing process. A method for estimating or obtaining appropriate item parameters prior to testing will be described
later.

\subsection{Algorithm for the adaptive online testing system}

Assuming that all item parameter values are known in advance, our system~\cite{LTLE2016c} employs the adaptive item selection algorithm described in Algorithm~\ref{AQS}. As shown in the algorithm, the first question is selected randomly, independent of the examinee's ability. The second item is also chosen at random but with consideration of the initial response: more difficult if the first question was answered correctly, and easier otherwise. From the third item onward, the selection becomes increasingly adaptive, reflecting the examinee's estimated ability.

\begin{algorithm}
\caption{Adaptive Question Selection}
\label{AQS}
\begin{algorithmic}
\medskip
\STATE Question 1:
\IF{questions $a=1, b=0$ exist}
\STATE select a question with $a=1, b=0$ at random
\ELSE
\STATE select a question with $ -0.25 \le b \le 0.25 $ at random
\ENDIF
\STATE provide the question to examinee
\STATE obtain the estimate of ability $ \hat{\theta}_1 $
\medskip
\STATE Question 2:
\IF{answer to question 1 is correct}
\IF{system can randomly select a question with $ 0.5 \le b \le 1.0 $ that has not been used before}
\STATE provide the question to examinee
\ELSE
\STATE search for the next appropriate question of $ b \ge 0 $ at random, and provide it to examinee
\ENDIF
\ELSE
\IF{system can randomly select a question with $ -1.0 \le b \le -0.5 $ that has not been used before}
\STATE provide the question to examinee
\ELSE
\STATE search for the next appropriate question of $ b \le 0 $ at random, and provide it to examinee
\ENDIF
\ENDIF
\STATE obtain the estimate of ability $ \hat{\theta}_2 $
\medskip
\STATE Questions $3, 4, \cdots, l, \cdots, k$:
\WHILE{$3 \le l \le k$}
\STATE select the unused question whose difficulty is closer to the most recently estimated ability $ \hat{\theta}_{l-1} $
\STATE provide the question to examinee
\STATE obtain the estimate of ability $ \hat{\theta}_l $
\ENDWHILE
\end{algorithmic}
\end{algorithm}

By administering $k$ items sequentially, we obtain $k$ ability estimates $\hat{\theta}_l$ for $l = 1, \dots, k$, as illustrated in Figures~\ref{figs:adaptivetest} and~\ref{figs:process}. Given the structure of the algorithm, the sequence of estimates is expected to converge toward the examinee's true ability $\tilde{\theta}$ as more items are answered. However, the initial estimates may deviate significantly from the true ability. As a result, the CAR is not necessarily centered at 0.5.
As previously discussed, when $\hat{\theta}_l < b$, where $b$ is the item difficulty, the probability of a correct answer is less than 0.5; conversely, when $\hat{\theta}_l > b$, the probability exceeds 0.5. In the next section, we investigate this phenomenon more meticulously using empirical data from the Linear Algebra module.

\medskip
It has been reported that, in order to accurately estimate a test-taker’s ability, a minimum of 10 questions is required~\cite{LTLE2016d, IJSCAI2017}, as using fewer items can introduce bias into the estimated ability. In the present study, however, the purpose of employing the adaptive system is not to obtain precise ability estimates, but rather to provide opportunities for practice and to promote understanding of fundamental mathematics without discouraging learners. Therefore, the number of questions was intentionally kept small, such as 5 or 7.
To enhance engagement, the system incorporates a “point system,” in which a test-taker’s accumulated points are displayed on a leaderboard if all questions in a given sequence are answered correctly; see Appendix~2. The scoring algorithm for this system is based on~\cite{point}. While this feature is designed to make studying more enjoyable, it may also introduce bias into both the ability estimates and the CAR values of the problem items. This paper examines the magnitude of such biases.

\medskip
To prevent potential misunderstandings, we define two statistics concerning the estimated abilities at each problem position in the adaptive test sequence. The first is the mean of all 
estimates $\hat{\theta}_j^{(l)}$ associated with problem $j$ across test-takers and positions, while the second is the mean of the final estimate $\hat{\theta}_j^{(k)}$ when problem $j$ appears as the final question in a sequence. These are formally defined as follows:
\begin{eqnarray} 
\label{eq:MeanEstimatesAbilityAll}
\hat{\mu}(\theta_j)_{\text {all}} = E(\hat{\theta}_j^{(l)} \mid j ),
\end{eqnarray} 
\begin{eqnarray} 
\label{eq:MeanEstimatesAbilityFinal}
\hat{\mu}(\theta_j)_{\text {final}} = E(\hat{\theta}_j^{(k)} \mid j ),
\end{eqnarray} 
where $E(\cdot \mid j)$ denotes the expectation conditioned on problem $j$ being presented at a particular position in the test sequence.
When asymptotic theory states that the CAR converges to 0.5 as $k \rightarrow \infty$, it implicitly assumes the use of $\hat{\mu}(\theta_j)_{\text{final}}$. However, when the number of items administered adaptively is not sufficiently large, $\hat{\mu}(\theta_j)_{\text{all}}$ may significantly differ from $\hat{\mu}(\theta_j)_{\text{final}}$. In what follows, we employ $\hat{\mu}(\theta_j)_{\text{all}}$ for direct comparison with observed CAR values.

For relevant adaptive online testing references, see~\cite{TALE2014c, TALE2014b, TALE2014a, BIC2016, IEE2018, PISM2018, IJLTLE2019a, IJLTLE2019b, Infomation2010, EeL2011, IPSJ2014, IJSCAI2017, LTLE2016d}. Also see references for systems that combine common IRT systems with online adaptive systems, e.g., \cite{LTLE2016a, LTLE2016b, LTLE2016c, LTLE2017, LTLE2018c, LTLE2018a, LTLE2018b, IEE2018, IJLTLE2019c, IJLTLE2022, Access2023}.

\section{Linear Algebra Case}

If the adaptive testing system can provide a sufficiently large pool of items with a wide range of difficulty levels and can accurately estimate an examinee's ability for each item, then the CAR ultimately converges to 0.5. In order to present an appropriately matched question to an examinee, the system must be able to assess the difficulty levels of all available items and have a prior approximation of the examinee’s ability. This is because the system aims to select the item whose difficulty is closest to the examinee’s estimated ability.

\subsection{Estimation for item parameters}

\begin{figure}[htbp]
\begin{center}
\includegraphics[height=11.5cm]{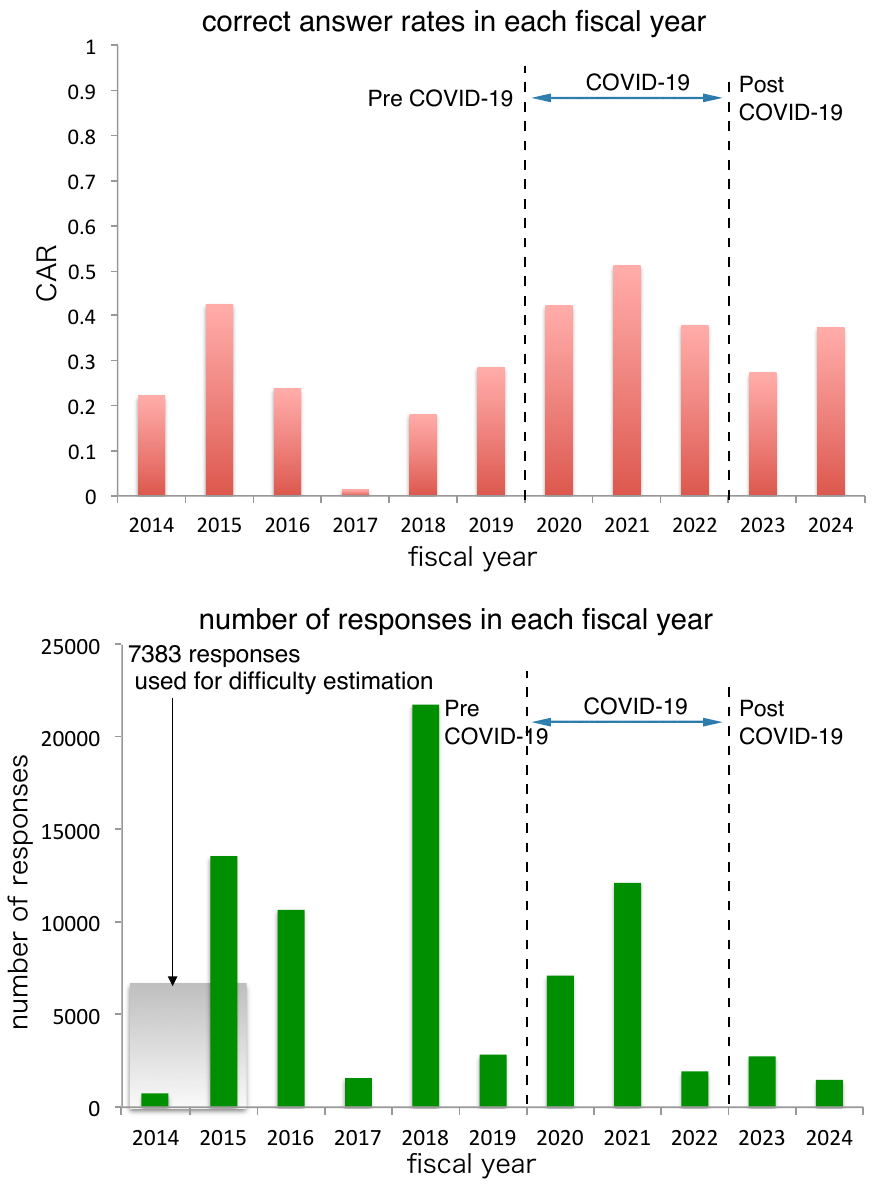}
\caption{Number of responses and CAR values for each fiscal year (Linear Algebra case).
As shown in the lower part of the figure, 7,383 responses were used to estimate item difficulty parameters. Since September 2015, these estimated parameters have been incorporated into the adaptive testing system.}
\label{figs:LACAR}
\end{center}
\end{figure}

However, item difficulties cannot be known without conducting actual testing. Therefore, as illustrated in the lower part of Figure \ref{figs:LACAR}, we designated the period from April 2014 to August 2015 as the difficulty estimation phase. The total number of responses collected during this period was 7,383.
Since the true difficulties of items were not available during this phase, we assigned default values of $a_j = 1$ and $b_j = 0$ to all items. This implies that questions were administered randomly, resulting in a response matrix containing missing values. Utilizing an estimation system capable of handling unobserved data, we obtained MAP estimates for item parameters from this incomplete response matrix~\cite{TALE2012, IPSJ2014}. 

Table \ref{table:ResponseMatrix} presents a subset of the estimated item parameters. Two types of estimates are shown: those obtained using all available items and those derived using only the items within each section. Given that our adaptive testing system is designed to function within individual sections, the latter approach is considered both appropriate and accurate.
As shown in Table \ref{table:ResponseMatrix}, there are not many differences between the two sets of estimated item parameters. As previously noted, our adaptive system operates within a single section. To ensure stable performance, each section should contain at least 40 to 50 questions, assuming that examinees answer approximately 5 to 10 consecutive questions in a single session.

\begin{table}[htbp]
\caption{Some estimated item parameters with incomplete response matrix.
\label{table:ResponseMatrix}}
\begin{center}
\begin{tabular}{c|c|cc|cc}
\hline
&& using all items && using section items & \\
\hline
section id & item id & item parameter && item parameter & \\
 &  & $a_j$ & $b_j$ & $a_j$ & $b_j$\\
\hline
30 & 715 & $1.061$ & $-0.685$ & $0.915$ & $-0.707$ \\
30 & 716 & $0.716$ & $0.382$ & $1.043$ & $0.561$ \\
30 & 717 & $0.835$ & $-0.764$ & $0.923$ & $-0.416$ \\
30 & 718 & $0.735$ & $0.425$ & $0.798$ & $0.568$ \\
30 & 719 & $0.905$ & $-0.417$ & $0.876$ & $-0.282$ \\
\vdots&\vdots& \vdots &\vdots&\vdots&\vdots \\
\hline
\end{tabular}
\end{center}
\end{table}

The upper part of Figure \ref{figs:LACAR} shows the CAR values for each fiscal year. During the COVID-19 pandemic, the CAR values appear slightly higher, though not significantly.

\begin{figure}[htbp]
\begin{center}
\includegraphics[height=6cm]{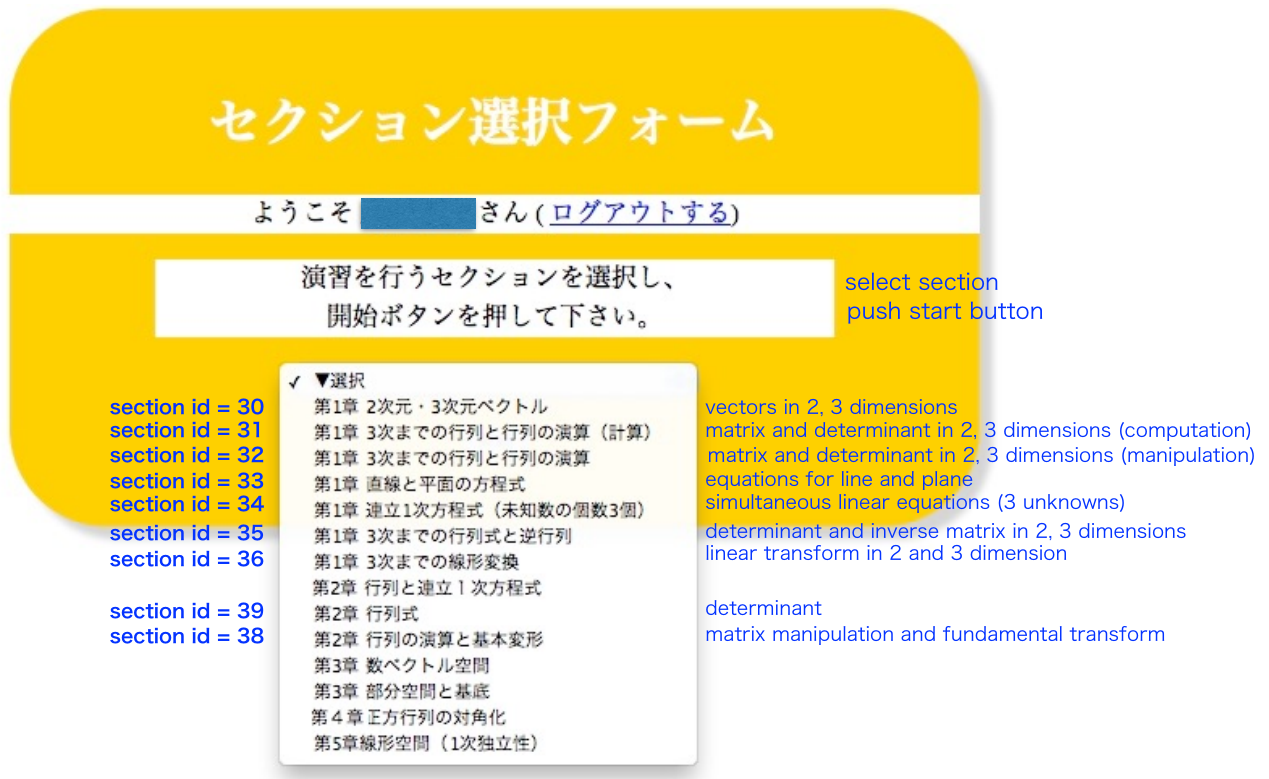}
\caption{Selecting a section in the Linear Algebra online adaptive system.}
\label{figs:sections}
\end{center}
\end{figure}

\medskip
As depicted in Figure \ref{figs:sections}, the Linear Algebra subject includes a large number of sections. Therefore, the total number of items, 860, as shown in Figure \ref{figs:ResponsesToSubjects}, is not particularly large. After logging in, an examinee selects a section, and the system then administers suitable questions selected from the pre-defined pool within that section.

\subsection{Analytics of adaptive online testing in practice}

Here, we describe the analytics of our adaptive online system, which has been in operation for over ten years, incorporating the proposed Algorithm~\ref{AQS}. Specifically, for each item, we can collect the test-taker's estimated ability $\hat{\theta}_j^{(l)}$ immediately after the item question is answered. By averaging these estimates, we obtain the mean estimated ability $\hat{\mu}(\theta_j)_{\text {all}}$ for each item. This value is then compared with the pre-obtained item difficulty ${b}_j$ and the corresponding CAR value to assess the effectiveness of the adaptive algorithm.

\begin{figure}[htbp]
\begin{center}
\includegraphics[height=8.5cm]{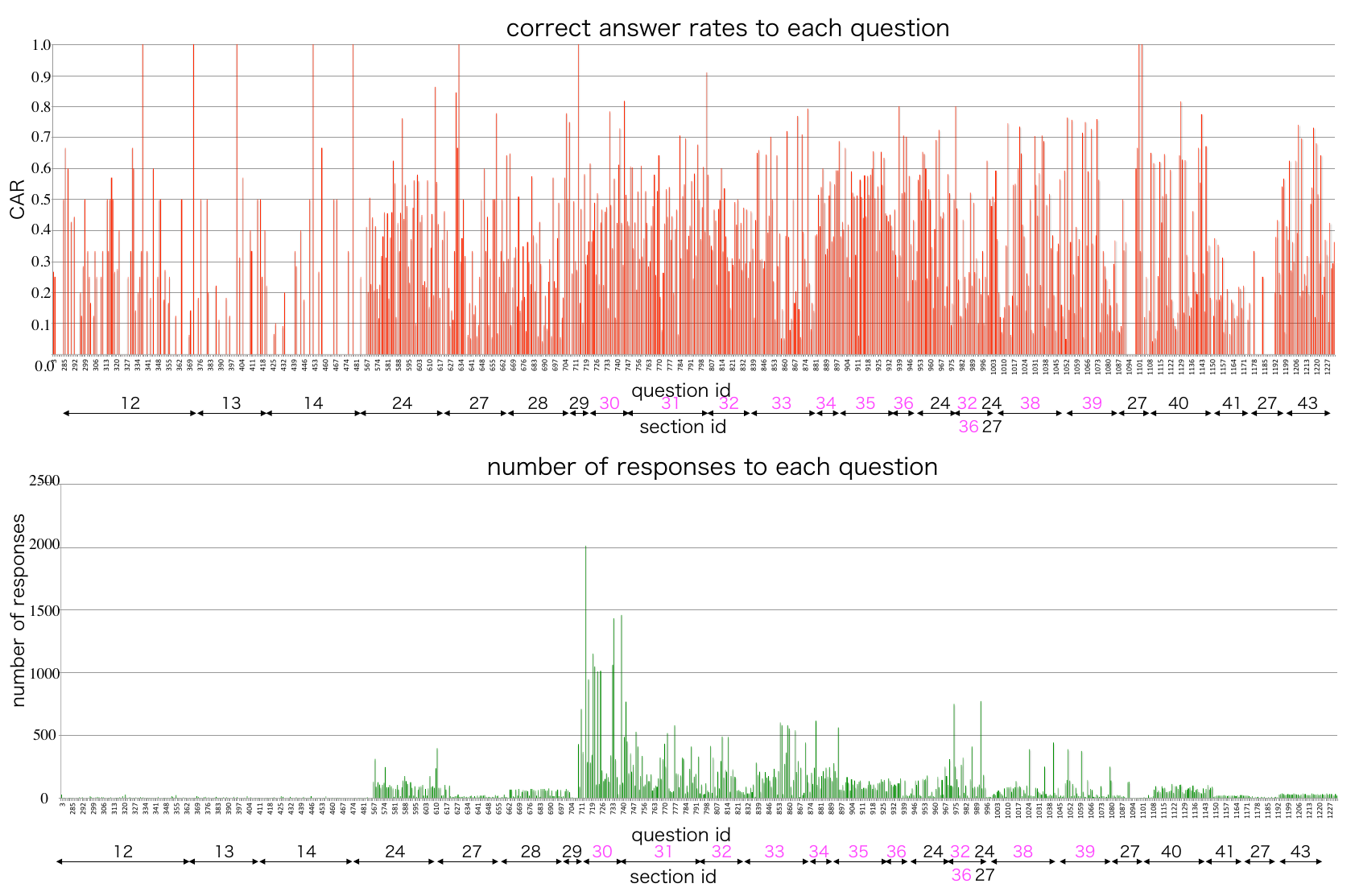}
\caption{Number of responses and CAR values for each problem item (Linear Algebra case).}
\label{figs:CARtoEachQuestion}
\end{center}
\end{figure}

Figure~\ref{figs:CARtoEachQuestion} presents the number of responses and the CAR values for each item. Sections 30 through 39, particularly sections 30, 31, 32, 33, 34, 35, 36, 38, and 39, exhibit frequent usage and considerable variation in CAR values. Therefore, we focus our analysis on these sections, examining the relationship among the CAR for item $j$, the mean estimated ability $\hat{\mu}(\theta_j)_{\text{all}}$, and the item difficulty $b_j$.

\subsubsection{Relation between item difficulty and mean estimated ability}

We first examine the relationship between item difficulty $b_j$ and the mean estimated ability for item $j$, $\hat{\mu}(\theta_j)_{\text{all}}$.

Table~\ref{table:CorrelationOfCoefficient} shows correlation coefficient between difficulty and mean estimated ability for each section.
As shown in the table,
a relatively strong correlation is observed between $b_j$ and $\hat{\mu}(\theta_j)_{\text{all}}$, 
in sections 30 through 35.
In contrast, sections 36 through 39 exhibit weaker correlations overall. 

\begin{figure}[htbp]
\begin{center}
\includegraphics[height=9cm]{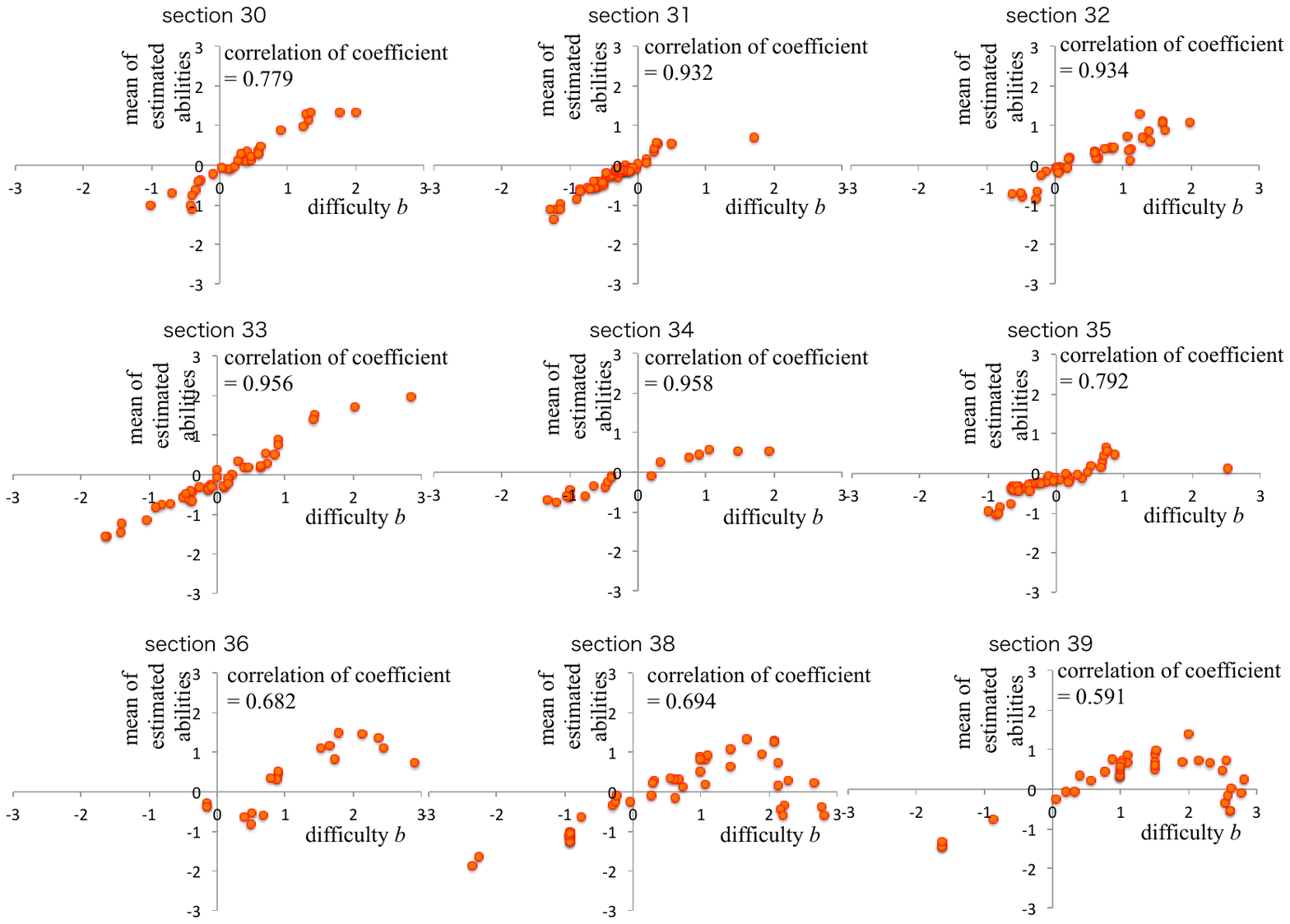}
\caption{Relationship between item difficulty $b_j$ and mean estimated ability $\hat{\mu}(\theta_j)_{\text {all}}$ for each section (Linear Algebra).}
\label{figs:bVStheta}
\end{center}
\end{figure}

Let’s take a look at Figure \ref{figs:bVStheta}, which illustrates the relationship between item difficulty $b_j$ and the mean estimated ability $\hat{\mu}(\theta_j)_{\text{all}}$ for each section.
Since a test taker’s ability is evaluated immediately after responding to the most recently answered item, the resulting estimate is directly influenced by the difficulty of that item.
Considering this, the items in sections 30 through 35 appear to be appropriately matched to the ability levels of the test takers.
In contrast, the weaker correlations observed in sections 36 through 39 suggest that these sections may contain many difficult items. Specifically, while the correlations for less difficult items within sections 36 through 39 remain relatively strong, similar to those in sections 30 to 35, for more difficult items, the mean estimated ability $\hat{\mu}(\theta_j)_{\text{all}}$ fluctuates around zero. This pattern suggests that test takers may be responding at random due to the increased complexity of the problems. This may be because the latter sections include more conceptually challenging problems rather than purely computational exercises.

\begin{table}[htbp]
\caption{Correlation coefficient between difficulty and mean estimated ability.
\label{table:CorrelationOfCoefficient}}
\begin{center}
\begin{tabular}{c|c}
\hline
section & correlation of coefficient \\
\hline
30 & 0.779 \\
31 & 0.932 \\
32 & 0.934 \\
33 & 0.956 \\
34 & 0.958 \\
35 & 0.792 \\
36 & 0.682 \\
38 & 0.694 \\
39 & 0.591 \\
\hline
\end{tabular}
\end{center}
\end{table}

Overall, the observed relationship between item difficulty and mean estimated ability indicates that the proposed algorithm is effectively performing the intended adaptive functionality of the online testing system.

\subsubsection{Relation between the number of correct answers and the number of responses}

In Figure \ref{figs:CARtoEachQuestion}, we previously examined CAR values without considering the number of responses. In this section, we investigate the CAR values for each item while accounting for the number of responses.

\begin{figure}[htbp]
\begin{center}
\includegraphics[height=9cm]{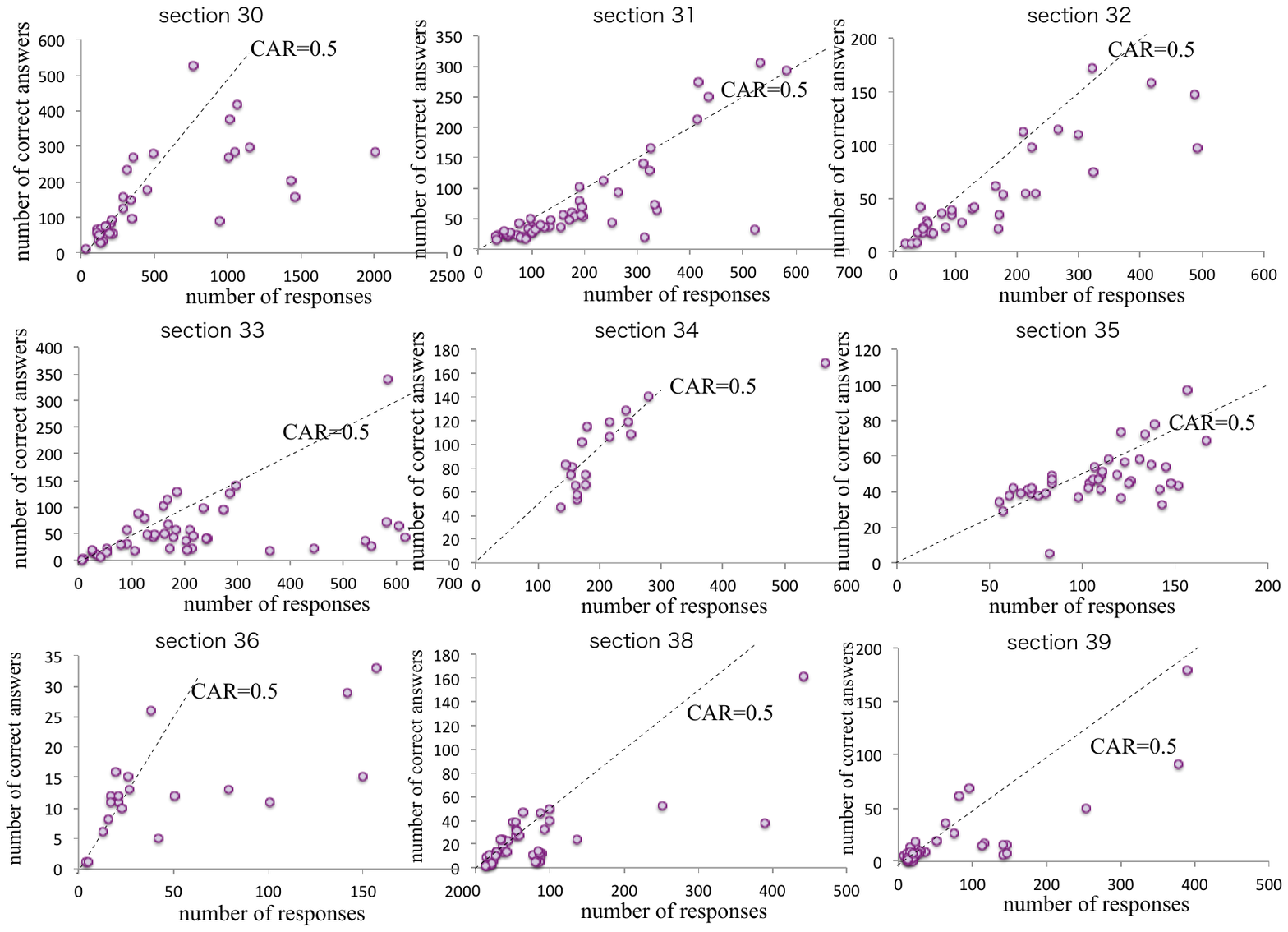}
\caption{Relationship between the number of responses and the number of correct answers for each section (Linear Algebra).}
\label{figs:CAR}
\end{center}
\end{figure}

Figure \ref{figs:CAR} illustrates the relationship between the number of responses and the number of correct answers for each section. In each subplot corresponding to a section ID, a dotted line represents a CAR value of 0.5. The prevalence of data points below this line indicates that the CAR value frequently falls below 0.5, which is consistent with the findings in Figure \ref{figs:CARtoEachQuestion}. Table \ref{table:EachCAR}, corresponding to Figure \ref{figs:CAR}, lists the CAR values for each problem item. Except for Section 34, the CAR values in almost all sections are below 0.5. Given that the number of responses is a dominant factor in determining the mean CAR, this result suggests that the CAR values are significantly lower than 0.5 in most sections, excluding Section 34.
This observation is particularly noteworthy: although the CAR values are consistently below 0.5, the system’s adaptive behavior remains effective, as previously discussed.

\begin{table}[htbp]
\caption{CAR values for each problem item.
\label{table:EachCAR}}
\begin{center}
\begin{tabular}{c|c}
\hline
section & CAR value \\
\hline
30 & 0.307 \\
31 & 0.377 \\
32 & 0.338 \\
33 & 0.257 \\
34 & 0.449 \\
35 & 0.442 \\
36 & 0.268 \\
38 & 0.299 \\
39 & 0.286 \\
\hline
\end{tabular}
\end{center}
\end{table}

\subsubsection{Relation between mean estimated ability and CAR}

\begin{figure}[htbp]
\begin{center}
\includegraphics[height=9cm]{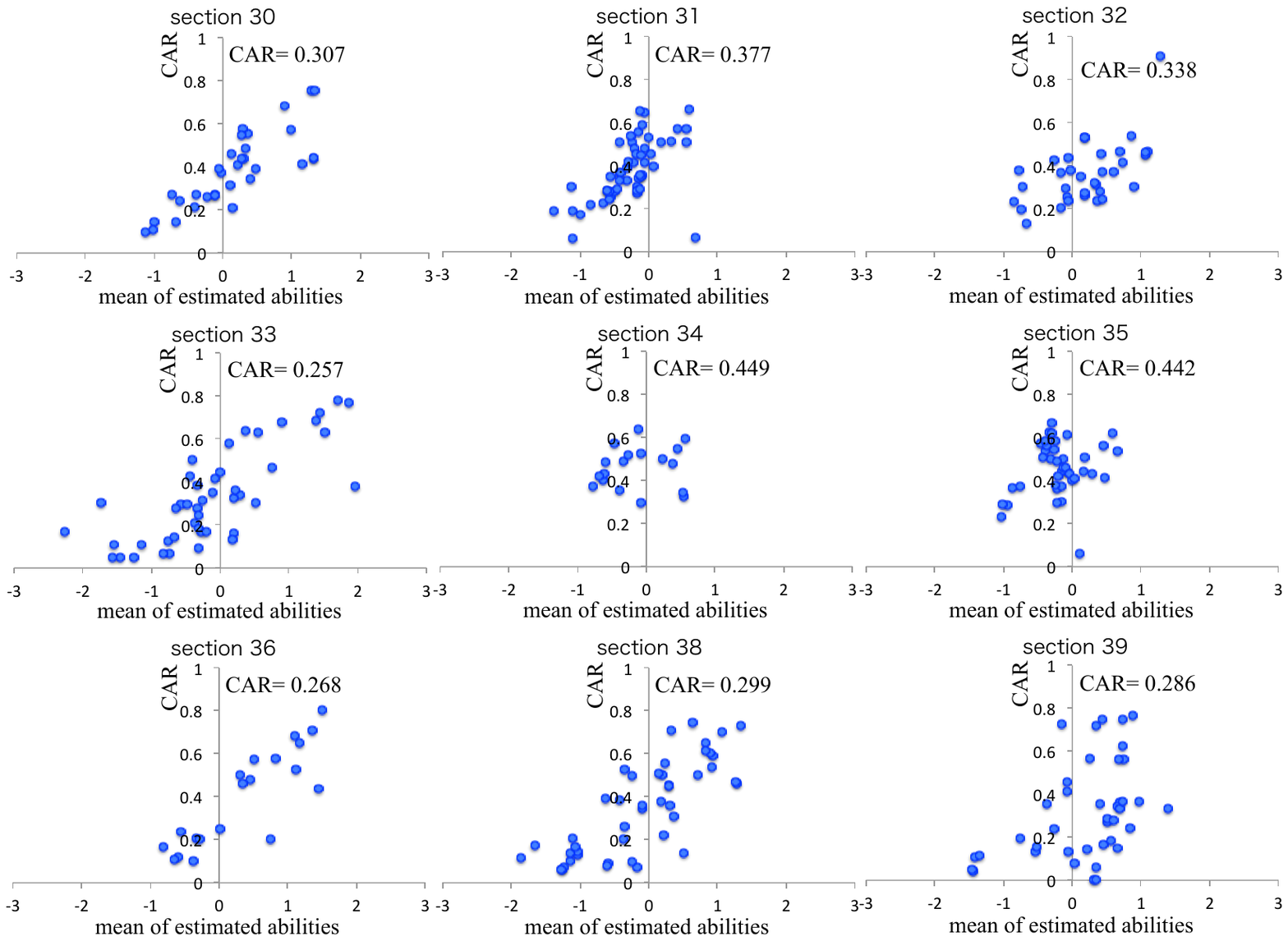}
\caption{Relationship between mean estimated ability $\hat{\mu}(\theta_j)_{\text {all}}$ and CAR value for each section (Linear Algebra).}
\label{figs:AbilityVsCAR}
\end{center}
\end{figure}

Next, Figure \ref{figs:AbilityVsCAR} presents the relationship between the mean estimated ability, $\hat{\mu}(\theta_j)_{\text{all}}$, and the CAR value. Note that the vertical axis does not represent the mean CAR value directly, since the number of responses for each point is not considered. If we compute a weighted mean using the number of responses as weights, the resulting value matches the CAR value for each section listed in Table \ref{table:EachCAR}.

Interestingly, with the exception of Sections 34 and 35, the relationship between the mean estimated ability $\hat{\mu}(\theta_j)_{\text{all}}$ and the CAR values appears to be approximately proportional. This is expected, as a higher estimated ability naturally corresponds to higher probabilities of correct responses. This observation further supports the effectiveness of the adaptive mechanism of the system, even when the number of consecutive items presented in a test is relatively small (e.g., five to seven items).

\subsubsection{Relation between mean estimated ability and problem-solving time}

So far, we have discussed how our proposed system maintains adaptiveness even with a relatively small number of items. It is well known that the larger the number of consecutive items, the more accurate the ability estimation tends to be. However, too many problems can cause test-takers to become fatigued and lose concentration. This suggests a trade-off between accuracy and the time required to complete the test.

\begin{figure}[htbp]
\begin{center}
\includegraphics[height=9cm]{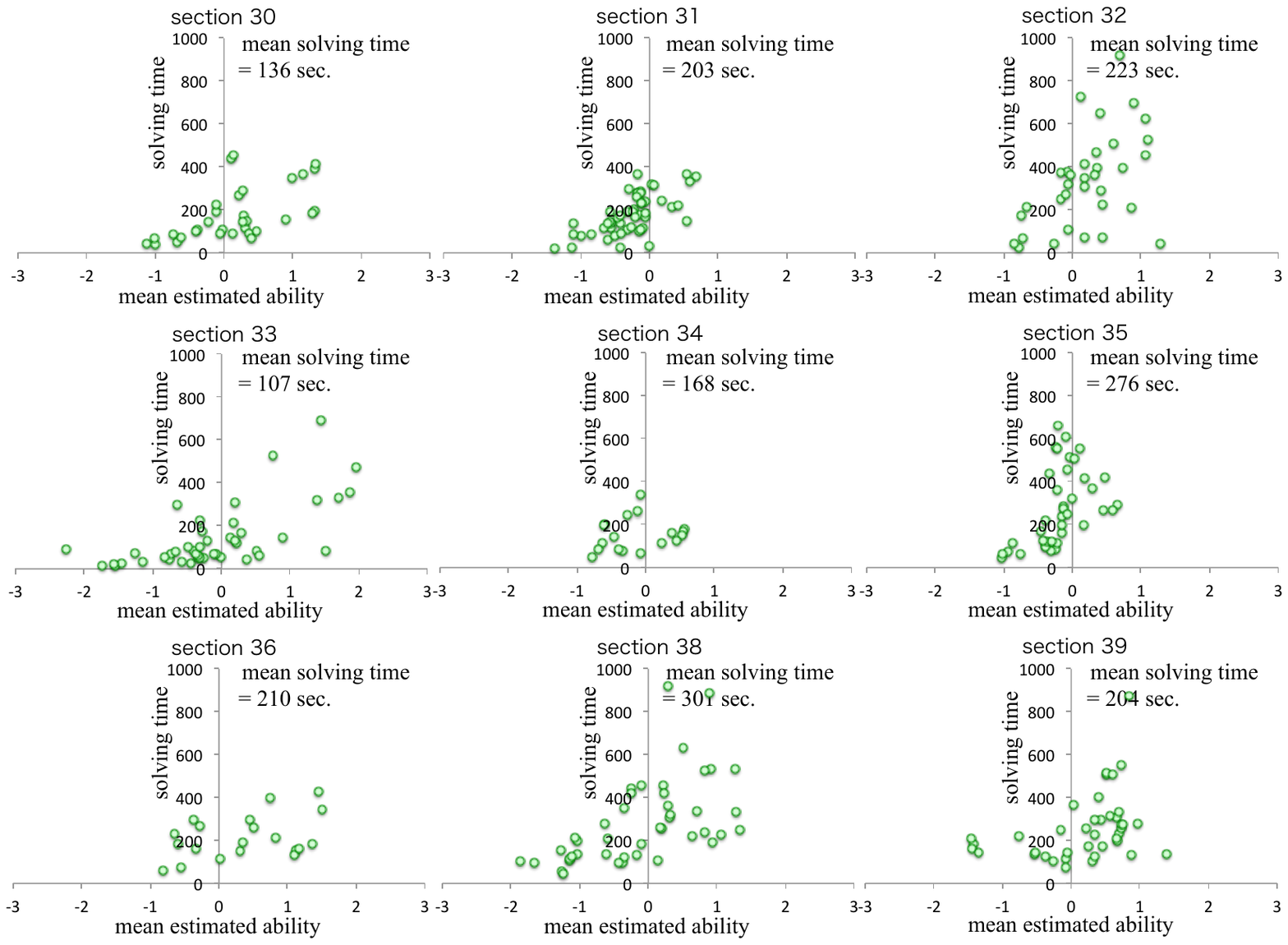}
\caption{Relationship between mean estimated ability and problem-solving time for each section (Linear Algebra).}
\label{figs:SolvingTime}
\end{center}
\end{figure}

Figure~\ref{figs:SolvingTime} 
shows the relationship between the mean estimated ability, $\hat{\mu}(\theta_j)_{\text{all}}$, and the mean problem-solving time for each section, which supports our decision to limit the number of items: more difficult problems tend to require more time, making overly long tests impractical. The mean problem-solving time for each item, $\hat{\mu}(T_j)_{\text{all}}$, is defined as
\begin{eqnarray} 
\label{eq:SolvingTime}
\hat{\mu}(T_j)_{\text{all}} = E(\hat{T}_j^{(l)} \mid j),
\end{eqnarray} 
where $T_j$ is the time duration in seconds for item $j$.

From the figure, we can see a slight positive relationship between the mean estimated ability and the solving time, although it is not particularly strong. This implies that more difficult problems generally require more time for a test-taker to tackle. For example, in section 38, the mean solving time is 301 seconds, or roughly 5 minutes. Assuming 5 such items, the total solving time would be around 25 minutes, which can be demanding in an online setting. In practice, 5 to 7 items may be a more reasonable range.

\subsubsection{Analytics summary}

Here, we summarize the findings regarding the interrelationship among the three key statistics, the CAR for item $j$, the mean estimated ability $\hat{\mu}(\theta_j)_{\text{all}}$, and the item difficulty $b_j$, discussed so far. Our analysis indicates that, in most cases, the CAR values in the adaptive testing system do not converge to the theoretical asymptotic value of 0.5. This is primarily because the system does not permit test takers to continuously attempt a sufficient number of items to allow the item difficulty and the test-taker's ability to align naturally over time.
At first glance, this may appear to undermine the adaptivity of the system. However, a more nuanced analysis, one that considers both the initially assigned difficulty levels and the mean estimated abilities obtained immediately after each response, reveals that the system does, in fact, exhibit adaptive behavior.

From a practical perspective, if the number of test items is increased excessively to improve estimation accuracy, it may lead to test taker fatigue or a decline in motivation to engage with the system. Our online platform places priority on promoting learner motivation and creating an enjoyable learning experience. Therefore, the fact that the CAR values do not approach 0.5 should not necessarily be viewed as a shortcoming.

Further research is warranted to determine the optimal number of test items that balances estimation accuracy with user engagement. Nonetheless, the current configuration, offering sequences of 5 to 7 items, appears to strike a reasonable balance, supporting both adaptivity and continued test participation in a remote learning environment.

\section{Concluding Remarks}

Adaptive online testing can assess the proficiency of examinees efficiently by dynamically adjusting test questions based on their performance. To achieve this, questions should be selected so that their difficulty matches the estimated ability at each testing time. This indicates that the probability of a correct answer when solving a problem ultimately approaches 0.5. However, in a real-world system, this probability will not reach 0.5 unless the number of given questions is large. It may seem that the adaptive feature is not functioning. However, if we focus on the relationship between the difficulty of the questions and the test taker's estimated ability at each time, we can see that the adaptive feature is functioning even when the number of questions is small, for example, 5 or 7. Similar findings can be obtained by using the relationship between the test taker's estimated ability and the correct answer rate. This proposed system is designed to accompany university mathematics textbooks with a system of online practice exercises.

\section*{Appendix}

{\bf Appendix 1: CAR values during the pre-COVID, COVID, and post-COVID periods}

\begin{figure}[htbp]
\begin{center}
\includegraphics[height=8cm]{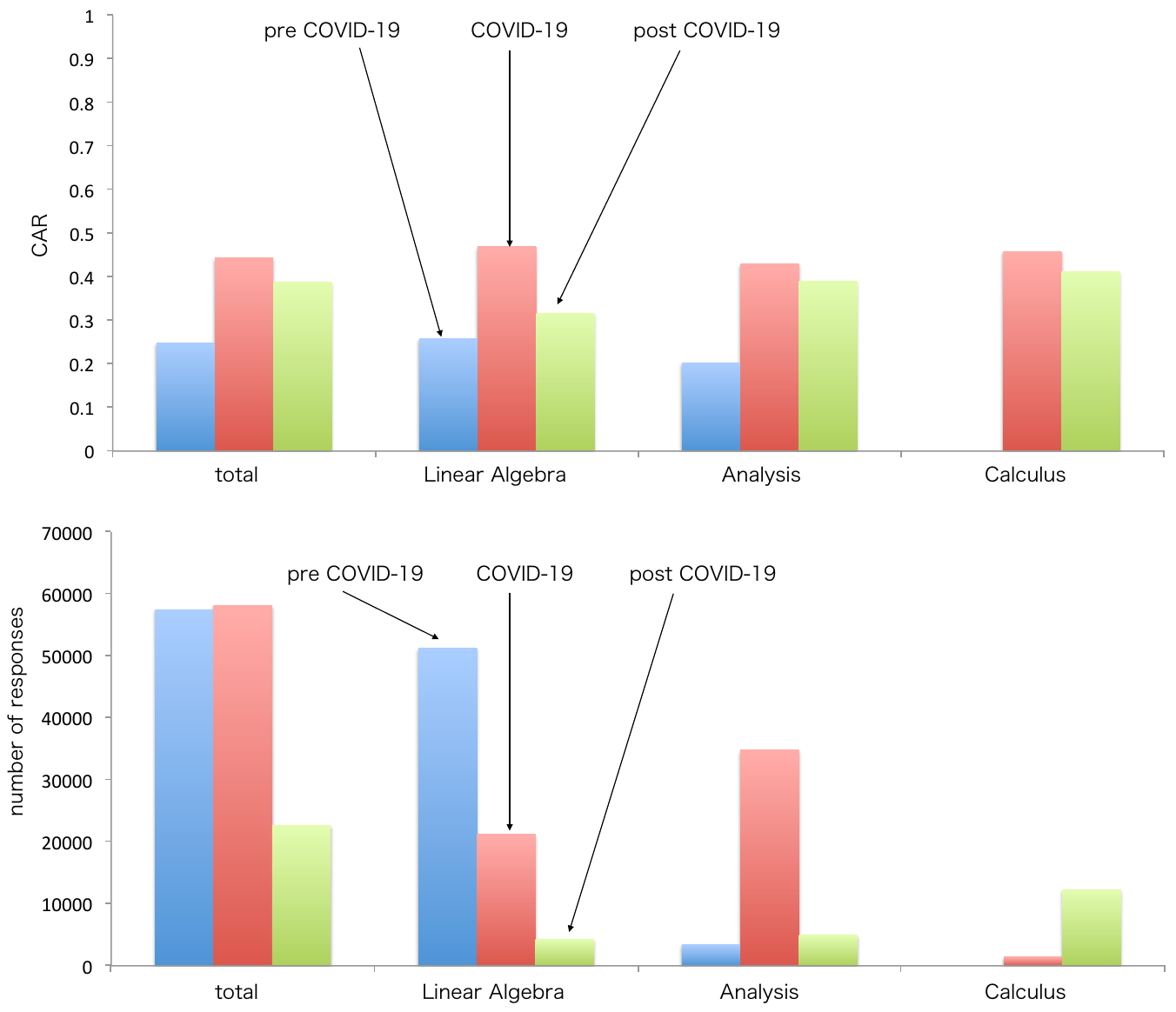}
\caption{CAR values and numbers of responses during the pre-COVID, COVID, and post-COVID periods (total, and for Linear Algebra, Analysis, and Calculus).}
\label{figs:CARofCOVID-19}
\end{center}
\end{figure}

\medskip
It is anticipated that test takers’ concerns about the system were heightened during the COVID-19 period compared to other times. Table \ref{table:CARofCOVID-19} presents the observed CAR values and their standard deviations for each subject. Since the standard deviations are notably small relative to the CAR values, the differences between the CAR values during the COVID-19 period and those in other periods can be considered statistically significant.
However, because the number of accesses to the subjects Probability and Statistics, Physics, and ODE was substantially lower than for other subjects, these were excluded from further analysis. Figure \ref{figs:CARofCOVID-19} (upper part) illustrates the CAR values across three distinct periods. The results suggest that, as expected, test takers engaged in more intensive study during the COVID-19 period than in other periods.
As shown in the lower part of Figure \ref{figs:CARofCOVID-19}, this conclusion is supported by the consistently higher CAR values for these subjects during the COVID-19 period, regardless of the absolute number of accesses in either period.

\begin{table}[htbp]
\caption{CAR values in pre-COVID, COVID, post-COVID periods.
\label{table:CARofCOVID-19}}
\begin{center}
\begin{tabular}{c|cccccc}
\hline
subject & pre & COVID & post & \\
             & observed (s.d.) & observed (s.d.) & observed (s.d.) \\
\hline
total & 0.249  (0.002) & 0.444  (0.002) & 0.388  (0.003)\\
Linear Algebra & 0.259  (0.002) & 0.470  (0.003) & 0.316  (0.007)\\
Probability \& Statics & 0.093  (0.007) & 0.315  (0.042) & 0.790  (0.041)\\
Analysis & 0.202  (0.007) & 0.430  (0.003) & 0.389  (0.007)\\
Physics & 0.183  (0.012) & 0.349  (0.033) & 0.489  (0.023)\\
Calculus I \& II &  & 0.458  (0.013) & 0.411  (0.004)\\
ODE &  & 0.225  (0.024) & 0.222  (0.019)\\
\hline
\end{tabular}
\center{s.d.: standard deviation}
\end{center}
\end{table}

\noindent
{\bf Appendix 2: Point system to make examinees fun in accessing the online system}

\begin{figure}[htbp]
\begin{center}
\includegraphics[height=6cm]{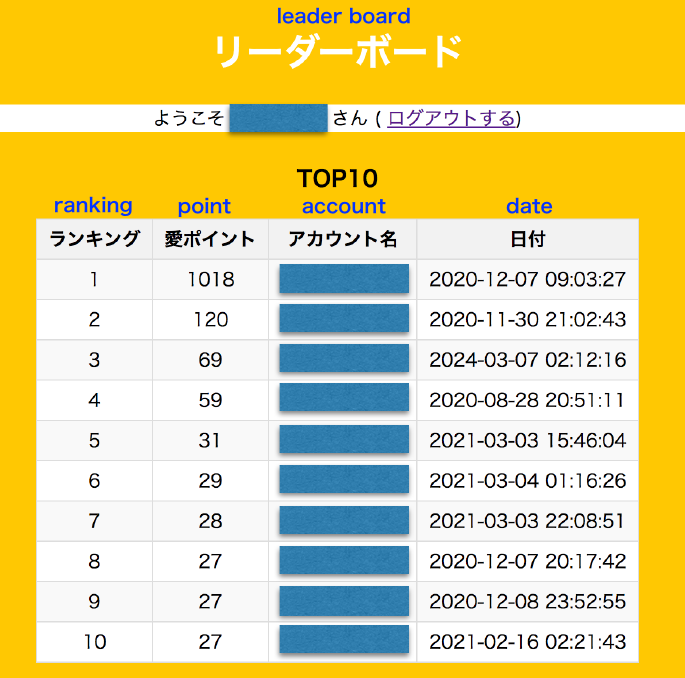}
\caption{Leader board presenting top 10 points (Analysis case).}
\label{figs:Leaderboard}
\end{center}
\end{figure}

\medskip
Figure \ref{figs:Leaderboard} illustrates the leaderboard of the “point system” incorporated into the online adaptive testing platform. When a test taker successfully solves all problems sequentially provided by the adaptive system, the point system awards additional points to their accumulated total. If the test taker continues to answer all questions correctly, their points increase exponentially, which may enhance test-taking motivation and engagement.
Table \ref{table:Leaderboard} summarizes the highest point totals and corresponding CAR values for each subject. With the exception of Probability and Statistics, subjects with higher CAR values tend to have top scorers with greater accumulated points.

\begin{table}[htbp]
\caption{Top point and CAR for each subject.
\label{table:Leaderboard}}
\begin{center}
\begin{tabular}{c|c|c}
\hline
subject & top points & CAR\\
\hline
Linear Algebra & 15 & 0.320\\
Probability \& Statistics & 58 & 0.144\\
Analysis & 1018 & 0.407\\
Physics & 6 & 0.285\\
Calculus I \& II & 120 & 0.416\\
ODE & 1 & 0.223\\
\hline
\end{tabular}
\end{center}
\end{table}

\section*{Acknowledgment}

The author would like to thank Drs. 
T. Katsura, Y. Sato, Y. Ikeda, 
T. Fujino, 
Y. Okazaki, T. Okayama, N. Saito, M. Tagami, M. Hirokado,
T. Osawa, S. Kuwata, K. Tanaka, M. Fujiwara, S. Otabe,
M. Takatou, Y. Yamauchi, T. Koyama, and
T. Wakasa
for providing the online problem items. 
He would also like to thank Messrs.  
T. Kuwahata, T. Sakumura, Y. Tsukihara, Y. Aizawa, N. Tabuchi, Y. Tokusada, K. Noguchi,
M. Kuwahara, and K. Nishi
for their help in building the online system.

This work was supported by 
JSPS KAKENHI Grant Numbers 23650543 and 17H01842.

\end{document}